\documentclass[apjl]{emulateapj}

\newcommand{\ctbd}[1]{}

\newcommand{\lc}{light-curve}
\newcommand{\lcs}{light-curves}

\newcommand{\cfa}{Harvard-Smithsonian Center for Astrophysics (CfA)}

\newcommand{\pxs}{\ensuremath{\rm \arcsec/pix}}

\newcommand{\ccdsize}[1]{\ensuremath{\rm #1\times\rm#1}}

\newcommand{\Ks}{\ensuremath{\rm K_S}}

\newcommand{\gmc}{\ensuremath{\rm g\,cm^{-3}}}

\newcommand{\teff}{\ensuremath{T_{\rm eff}}}
\newcommand{\logg}{\ensuremath{\log{g}}}

\newcommand{\rsun}{\ensuremath{R_\sun}}
\newcommand{\msun}{\ensuremath{M_\sun}}

\newcommand{\teffsun}{\ensuremath{T_{eff,\sun}}}

\newcommand{\rstar}{\ensuremath{R_*}}
\newcommand{\mstar}{\ensuremath{M_*}}

\newcommand{\teffstar}{\ensuremath{T_{eff,*}}}

\newcommand{\rpl}{\ensuremath{R_{\rm P}}}

\newcommand{\ipl}{\ensuremath{i_{\rm P}}}

\newcommand{\rjup}{\ensuremath{R_{\rm J}}}
\newcommand{\mjup}{\ensuremath{M_{\rm J}}}

\newcommand{\prog}[1]{\textsc{\lowercase{#1}}}
\newcommand{\iraf}{\prog{iraf}}

\newcommand{\daophot}{\prog{daophot}}

\newcommand{\figr}[1]{Fig.~\ref{fig:#1}}
\newcommand{\secr}[1]{\S\ref{sec:#1}}
\newcommand{\eqr}[1]{Eq.~\ref{eq:#1}}

\newcommand{\tabr}[1]{\mbox{Table~\ref{tab:#1}}}

\defcitealias{bouchy05}{B05}

\newcommand{\hds}{\mbox{HD 189733}}
\newcommand{\hdsb}{\mbox{HD 189733b}}
\newcommand{\hdsB}{\mbox{HD 189733B}}

\newcommand{\hdstb}{\mbox{HD 209458b}}

\shorttitle{Refined parameters of the HD 189733 system}
\shortauthors{Bakos et al.}

\begin{document}
\title{Refined parameters of the planet orbiting HD 189733
}
\author{
G.~\'A.~Bakos\altaffilmark{1,2}, 
H.~Knutson\altaffilmark{1},
F.~Pont\altaffilmark{5},
C.~Moutou\altaffilmark{3},
D.~Charbonneau\altaffilmark{1},
A.~Shporer\altaffilmark{8},
F.~Bouchy\altaffilmark{4,10},
M.~Everett\altaffilmark{6},  
C.~Hergenrother\altaffilmark{7},
D.~W.~Latham\altaffilmark{1},
M.~Mayor\altaffilmark{5},
T.~Mazeh\altaffilmark{8},
R.~W.~Noyes\altaffilmark{1},
D.~Queloz\altaffilmark{5},
A.~P\'al\altaffilmark{9,1} and
S.~Udry\altaffilmark{5}
}
\email{gbakos@cfa.harvard.edu}
\altaffiltext{1}{\cfa,
	60 Garden Street, Cambridge, MA 02138, USA}
\altaffiltext{2}{Hubble Fellow}
\altaffiltext{3}{Laboratoire d'Astrophysique de Marseille,
	Traverse du Siphon, 13013 Marseille, France}
\altaffiltext{4}{Observatoire de Haute Provence,
	04870 St Michel l'Observatoire, France}
\altaffiltext{5}{Observatoire de Gen\`eve,
	51 ch.~des Maillettes, 1290 Sauverny, Switzerland}
\altaffiltext{6}{Planetary Science Institute,
	Fort Lowell Rd.,Tucson, AZ 85719, USA}
\altaffiltext{7}{Department of Planetary Sciences \&
	Lunar and Planetary Laboratory,
	The University of Arizona,
	1629 E.~University Blvd.~Tucson, AZ 85721, USA}
\altaffiltext{8}{Wise Observatory, Tel Aviv University, Tel Aviv,
	Israel 69978}
\altaffiltext{9}{E\"otv\"os Lor\'and University, Department of
	Astronomy, H-1518 Budapest, Pf.~32., Hungary}
\altaffiltext{10}{Institut d'Astrophysique de Paris, 98bis Bd Arago, 
	75014 Paris, France}
\setcounter{footnote}{10}

\begin{abstract}

We report on the BVRI multi-band follow-up photometry of
the transiting extrasolar planet \hdsb. We revise the
transit parameters and find planetary radius 
$\rpl = 1.154\pm0.032\rjup$ and inclination 
$\ipl = 85.79\pm0.24\arcdeg$. The new density 
($\rm \sim 1g\,cm^{-3}$) is significantly higher than the
former estimate 
($\rm \sim 0.75g\,cm^{-3}$); this shows
that from the current sample of 9 transiting planets, only
{\mbox HD 209458} (and possibly OGLE-10b) have anomalously
large radii and low densities. We note that due to the
proximity of the parent star, \hdsb\ currently has one of
the most precise radius determinations among extrasolar
planets.  We calculate new ephemerides:
$P = 2.218573\pm0.000020$ days,
$T_0 = 2453629.39420\pm0.00024$ (HJD), and 
estimate the timing offsets of the 11 distinct transits
with respect to the predictions of a constant orbital
period, which can be used to reveal the presence of
additional planets in the system.

\end{abstract}

\keywords{
stars: individual: \hds\ \---
planetary systems
}

\section{Introduction}

\hds\ is one of nine currently known main sequence stars orbited by a
transiting giant planet. The system is of exceptional interest because
it is the closest known transiting planet (D = 19.3pc), and thus is
amenable to a host of follow-up observations.  The discovery paper
by \citet{bouchy05} (hereafter \citetalias{bouchy05}) derived the key
physical characteristics of the planet, namely its mass ($1.15\pm
0.04\mjup$) and radius ($1.26\pm0.03\rjup$), based on radial
velocity observations of the star made with the ELODIE spectrograph at
the 1.93m telescope at Observatoire de Haute Provence (OHP), together
with photometric measurements of one complete and two partial transits
made with the 1.2m telescope also at OHP\@. With these parameters \hdsb\
had a large radius comparable to \hdstb\ \citep{laughlin05}, and a
density roughly equal to that of Saturn ($\rho\sim0.7\gmc$).

Determining precise radii of extrasolar planets in addition to their
mass is an important focus of exoplanet research
\citep[see~e.g.][]{bouchy04,torres04}, because the mean density
of the planets can shed light on their internal structure and
evolution. According to \citet{baraffe05}, the radii of all known
extrasolar planets are broadly consistent with models, except for
\hdstb. This planet with its large radius and low density
($\rho\sim0.33\gmc$) has attracted considerable interest, and various
mechanisms involving heat deposition beneath the surface
have been suggested \citep[][ and references therein]{laughlin05}.
An additional motivation for obtaining accurate planetary radii is
proper interpretation of follow-up data, notably secondary eclipse
and reflected light observations.  This is of particular relevance to
\hdsb, which has been recently observed by the Spitzer Space Telescope
\citep[][]{deming06}, and
where the brightness temperature depends on the radius ratio of the
planet to the star.

Both by extending the current, very limited sample of transiting
exoplanets, and by precise determination of the physical parameters
it will become possible to refine theoretical models and decide which
planets are ``typical''. Close-by, bright stars, such as the host star
of \hds\ are essential in this undertaking. The OGLE project
\citep{udalski02a,udalski02b,udalski02c} and follow-up observations
\citep[e.g.][]{konacki04,moutou04,pont05a} made a pivotal contribution
to the current sample by the discovery of
more than half of the known transiting planets. Follow-up observations,
however, are cumbersome due to the faintness of the targets, and
required the largest available telescopes. The typical errors of mass
and radius for these host stars are $\sim0.06\msun$ and $\sim 0.15\rsun$, 
and the corresponding errors in planetary parameters are 
$\sim 0.13\mjup$ and $\sim 0.12\rjup$, respectively. 
However, for planets orbiting bright stars in the solar neighborhood,
errors at the level of a few percent can be reached for both the mass
and radius.

In this paper we report a number of follow-up photometric measurements
of \hds\ using six telescopes spaced around the world. Together with
the original OHP photometry, we use these measurements to determine
revised values for the transit parameters, and give new ephemerides.
First we describe the follow-on photometry in detail (\secr{obsred}),
followed by the modeling which leads to the revised estimate of the
planetary radius (\secr{physpar}), and we conclude the paper in
\secr{conc}.

\notetoeditor{This is the original place of \tabr{inst}.}
\begin{deluxetable*}{lllrlllrrr}
\tabletypesize{\scriptsize}
\tablecaption{
Summary of instruments used in the observing campaign of
\hds.
\label{tab:inst}}
\tablewidth{0pt}
\tablehead{
	\colhead{Site} &
	\colhead{Longitude} &
	\colhead{Latitude} &
	\colhead{Alt.} &
	\colhead{Telescope} &
	\colhead{Diam.} &
	\colhead{Detector} &
	\colhead{Pxs} &
	\colhead{$T_{rd}$} &
	\colhead{FOV}\\
	& 
	& 
	& 
	\colhead{(meters)} &
	&
	\colhead{(meters)} & 
	&
	\colhead{(\pxs)} &
	\colhead{(sec)} &	
}
\startdata
       OHP & \phn05\arcdeg30\arcmin\ E & 43\arcdeg55\arcmin\ N & 650  &  OHP1.2 &  1.2 & SITe \ccdsize{1K}       & 0.69     &   90 & 11.77\arcmin \\
      FLWO & 110\arcdeg53\arcmin\ W    & 31\arcdeg41\arcmin\ N & 2350 & FLWO1.2 &  1.2 & Fairchild \ccdsize{4K}  & 0.34     &   12 &    23\arcmin \\
      FLWO & 110\arcdeg53\arcmin\ W    & 31\arcdeg41\arcmin\ N & 2345 &   HAT-5 & 0.11 & Thomson \ccdsize{2K}    & 14.0\phn &   10 &   8.2\arcdeg \\
      FLWO & 110\arcdeg53\arcmin\ W    & 31\arcdeg41\arcmin\ N & 2345 &   HAT-6 & 0.11 & Thomson \ccdsize{2K}    & 14.0\phn &   10 &   8.2\arcdeg \\
      FLWO & 110\arcdeg53\arcmin\ W    & 31\arcdeg41\arcmin\ N & 2345 &  TopHAT & 0.26 & Marconi \ccdsize{2K}    & 2.2\phn  &   40 &  1.29\arcdeg \\
 Mauna Kea & 155\arcdeg28\arcmin\ W    & 19\arcdeg49\arcmin\ N & 4163 &   HAT-9 & 0.11 & Thomson \ccdsize{2K}    & 14.0\phn &   10 &   8.2\arcdeg \\
      Wise & \phn34\arcdeg35\arcmin\ E & 30\arcdeg35\arcmin\ N & 875  & Wise1.0 &  1.0 & Tektronics \ccdsize{1K} & 0.7\phn  &   40 & 11.88\arcmin \\
\enddata
\end{deluxetable*}

\section{Observations and Data Reduction}
\label{sec:obsred}

We organized an extensive observing campaign with the goal of acquiring
multi-band photometric measurements of the transits of \hds\ caused by
the hot Jupiter companion. Including the discovery data of
\citetalias{bouchy05} that were obtained at OHP, altogether four sites
with seven telescopes contributed data to two full and eight partial
transits in Johnson B, V, R, I and Sloan r photometric bandpasses. The
sites and telescopes employed are spread in geographic longitude, which
facilitated gathering the large number (close to 3000) of individual
data spanning 2 months.

The following telescopes were involved in the photometric monitoring:
the 1m telescope at the Wise Observatory, Israel; the 1.2m telescope
at OHP; the 1.2m telescope
at Fred Lawrence Whipple Observatory (FLWO) of the Smithsonian
Astrophysical Observatory (SAO); the 0.11m HAT-5 and HAT-6 wide field
telescopes plus the 0.26m TopHAT telescope also at FLWO; and the 0.11m
HAT-9 telescope at the Submillimeter Array site at Mauna Kea, Hawaii.
An overview of the sites and telescopes is shown in \tabr{inst}.

A summary of the observations is shown in \tabr{obsstat}. The
telescopes are identified by the same names as in \tabr{inst}. The
transits have been numbered starting with the discovery data
$N_{tr}\equiv0$, and are identified later in the text using this number. 
In the following subsections we summarize the observations and
reductions that are specific to the sites or instruments.

\notetoeditor{This is the original place of \tabr{obsstat}.}
\begin{deluxetable*}{llllllccclr}
\tabletypesize{\scriptsize}
\tablecaption{
Summary of \hds\ observations.
\label{tab:obsstat}}
\tablewidth{0pt}
\tablehead{
	\colhead{Telescope} &
	\colhead{Filter} &
	\colhead{$N_{tr}$} &
	\colhead{Epoch} &
	\colhead{Date} &
	\colhead{Transit} &
	\colhead{Cond.} &
	\colhead{$\sigma_{OOT}$} &
	\colhead{$\sigma_{sys}$} &
	\colhead{Cad.} &
	\colhead{Ap[\arcsec]}\\
	\colhead{} &
	\colhead{} &
	\colhead{} &
	\colhead{} &
	\colhead{(UT)} &
	\colhead{} &
	\colhead{} &
	\colhead{(mmag)} &
	\colhead{(mmag)} &
	\colhead{(sec)} &
	\colhead{(\arcsec)}
}
\startdata
  OHP1.2 & B                  & 0    & 53629.4 & 2005-09-15 & {\tt OIBEO} & 5                  & 2.6  &  1.3 &   86 &   10 \\
 Wise1.0 & B                  & 4    & 53638.3 & 2005-09-24 & {\tt -IBE-} & 4                  & \nodata & \nodata &   42 &    5 \\
  OHP1.2 & $\rm R_c$          & 4    & 53638.3 & 2005-09-24 & {\tt --BEO} & 4                  & 3.0  &  1.2 &   95 &   10 \\
  OHP1.2 & $\rm R_c$          & 5    & 53640.5 & 2005-09-26 & {\tt OI---} & 3                  & 6.8  &  2.4 &   95 &   10 \\
 FLWO1.2 & r\tablenotemark{a} & 6    & 53642.7 & 2005-09-29 & {\tt OIBEO} & 4\tablenotemark{b} & 2.6  &  0.5 &   17 &   20 \\
   HAT-5 & $\rm I_c$          & 6    & 53642.7 & 2005-09-29 & {\tt OIBEO} & 4\tablenotemark{b} & 4.4  &  1.3 &  135 &   42 \\
   HAT-6 & $\rm I_c$          & 6    & 53642.7 & 2005-09-29 & {\tt OIBEO} & 4\tablenotemark{b} & 4.1  &  1.2 &  108 &   42 \\
  TopHAT & V                  & 6    & 53642.7 & 2005-09-29 & {\tt OIBEO} & 4\tablenotemark{b} & 4.6  &  3.0 &   70 &   10 \\
   HAT-9 & $\rm I_c$          & 7    & 53644.9 & 2005-10-01 & {\tt OIB--} & 4                  & 4.6  &  1.2 &   99 &   42 \\
   HAT-9 & $\rm I_c$          & 16   & 53664.9 & 2005-10-21 & {\tt OIB--} & 4                  & 4.3  & \nodata &  100 &   42 \\
  TopHAT & V                  & 19   & 53671.6 & 2005-10-28 & {\tt ---eO} & 4                  & 5.3  & \nodata &  106 &   10 \\
   HAT-5 & $\rm I_c$          & 19   & 53671.6 & 2005-10-28 & {\tt ---EO} & 4                  & 4.6  & \nodata &  103 &   42 \\
   HAT-5 & $\rm I_c$          & 20   & 53673.8 & 2005-10-30 & {\tt OIb--} & 5                  & 3.3  &  0.9 &   85 &   42 \\
 Wise1.0 & B                  & 22   & 53678.2 & 2005-11-03 & {\tt --bEO} & 3                  & 5.5  &  1.1 &   49 &   10 \\
  TopHAT & V                  & 24   & 53682.6 & 2005-11-08 & {\tt OIb--} & 2                  & 5.4  &  2.6 &  108 &   10 \\
   HAT-9 & $\rm I_c$          & 29   & 53693.7 & 2005-11-19 & {\tt -IBEO} & 5                  & 6.6  &  1.1 &   90 &   42 \\
\enddata
\tablecomments{
The table summarizes {\em all}\/ observations that were part of
the observing campaign described in this paper. Not all of them were
used for refining the ephemerides or parameters of the transit \--- see
\tabr{lc} and \tabr{trloc} for reference. {\bf $\mathbf{N_{tr}}$} shows the
number of transits since the discovery data. {\bf Epoch} and {\bf Date}
show the approximate time of mid-transit. The {\bf Transit} column
describes in a terse format which parts of the transits were observed;
Out-of-Transit (OOT) section before the transit, Ingress, Bottom,
Egress and OOT after the transit. Missing sections are indicated by
"\---". The {\bf Conditions} column indicates the photometric
conditions on the scale of 1 to 5, where 5 is absolute photometric, 4
is photometric most of the time with occasional cirrus/fog (relative
photometric), 3 stands for broken cirrus, and 2 for poor conditions.
Column {\bf $\sigma_{OOT}$} gives the rms of the OOT section
at the {\bf Cadence} shown in the next column. If the
transit was full, $\sigma_{OOT}$ was computed separately from the pre-
and post-transit data, and the smaller value is shown. Column {\bf
$\sigma_{sys}$} shows the estimated amplitude of systematics
(for details, see \secr{trfit}). {\bf Ap} shows
the aperture used in the photometry in arcseconds.
}
\tablenotetext{a}{Sloan r filter}
\tablenotetext{b}{Conditions were non-photometric before the
transit (on the initial part of the OOT), then they became photometric
for the entire duration of the transit, and deteriorated after
the transit at the end of the OOT.}
\end{deluxetable*}

\subsection{Observations by OHP 1.2m telescope}

These observations and their reduction were already described in
\citetalias{bouchy05}. Summarizing briefly, the 1.20m f/6 telescope
was used together with a \ccdsize{1K} 
back-illuminated CCD
having 0.69\pxs\ resolution. Typical exposure times were 6 seconds
long, followed by a 90 second readout. The images were slightly
defocused, with FWHM$\approx$2.8\arcsec.

Full-transit data were obtained in Johnson B-band under photometric
conditions for the $N_{tr}=0$ transit. This is shown by the ``OIBEO''
flag in \tabr{obsstat} indicating that the Out-of-transit part before
the Ingress, the Bottom, the Egress, and the Out-of-transit part after
egress all have been observed. This is an important part of the
combined dataset, as it is the only full transit seen in B-band.
In addition, partial transit data were obtained for the $N_{tr}=4$
event using Cousin's R-band filter ($R_C$) under acceptable photometric
conditions, and for the $N_{tr}=5$ event using the same filter under
non-photometric conditions.
The frames were subject to bias, dark and flatfield calibration
procedure followed by cosmic ray removal. Aperture photometry was
performed in an aperture of 9.6\arcsec\ using the \daophot\ 
\citep{stetson87} package.

The B-band \lc\ published in the \citetalias{bouchy05} discovery paper
used the single comparison star \mbox{HD 345459}. This
\lc\ suffered from a strong residual trend, as suggested by the
$\sim 0.01mag$ difference between the pre- and post-transit sections. 
This trend was probably a consequence of differential atmospheric
extinction, and was removed by a linear airmass correction,
bringing the two sections
to the same mean value. This ad-hoc correction, however, may have
introduced an error in the transit depth.
In this paper, we used six comparison stars in the field of view
(selected to have comparable relative flux to \hds\ before and after
transit). A reference \lc\ was built by co-adding the normalized flux
of all six stars, and was subtracted from the normalized \lc\ of \hds.
The new reduction shows a residual OOT slope 4.2 times smaller than in
the earlier reduction. The resulting transit depth in B-band is
decreased by about 20\% compared to the discovery data. This
illustrates the large contribution of photometric systematics that
must be accounted for in this kind of measurement. The R-band data set
is not as sensitive to the extinction effect as the B-band, hence the
selection of comparison stars has a minimal impact on the shape of the
transit curve. The B-band \lc\ is shown on panel 6 of
\figr{fulltr}, the R-band \lcs\ are exhibited in \figr{partr}.

\subsection{Observations by the FLWO 1.2m telescope}
\label{sec:flwo48obs}

We used the FLWO 1.2m telescope to observe the full transit of
$N_{tr}=6$
in Sloan r band. The detector was Keplercam
, which is a single chip \ccdsize{4K} 
CCD with 15\micron\ pixels that correspond to 0.34\arcsec\ on the sky.
The entire field-of-view is 23\arcmin. The chip is read out by 4
amplifiers, yielding a 12 second readout with the $2\times2$ binning we
used. The single-chip design, wide field-of-view, high sensitivity and
fast readout make this instrument well-suited for high-quality
photometry follow-up.

The target was deliberately defocused in order to allow longer exposure
times without saturating the pixels, and to smear out the inter-pixel
variations that may remain after flatfield calibration. The intrinsic
FWHM was $\sim2\arcsec$, which was defocused to $\sim10\arcsec$. While
conditions during the transit were photometric\footnote{This was
confirmed from the all-sky webcamera movies taken at the
Multiple-Mirror Telescope (MMT) that are archived on a nightly basis
(http://skycam.mmto.arizona.edu/)}, there were partial clouds before
and after. The focus setting was changed twice during the night; first
when the clouds cleared, and second, when the seeing improved. In both
cases the reason was to keep the signal level within the
linear response range of the CCD\@.
We used a large enough aperture that these focus changes did not affect
the photometry. All exposures were 5 seconds in length with 12 seconds
of readout and overhead time between exposures. We observed the target
in a single band so as to maximize the cadence, and to eliminate
flatfielding errors that may originate from the imperfect sub-pixel
re-positioning of the filter-wheel. Auto-guiding was used to further
minimize systematic errors that originate from the star drifting away
on the CCD chip and falling on pixels with different (and not perfectly
calibrated) characteristics.

\paragraph{Reduction and photometry}

All images were reduced in the same manner; applying overscan
correction, subtraction of the two-dimensional residual bias pattern
and correction for shutter effects.
Finally we flattened each image using a combined and normalized set of
twilight sky flat images.  There was a drift of only $\sim3\arcsec$ in
pointing during the night, so any large-scale flatfielding errors were
negligible.

To produce a transit light curve, we chose one image as an astrometric
reference and identified star centers for \hds\ and 23 other bright and
relatively uncrowded stars in the field.
We measured the flux of each star around a fixed pixel center derived
from an astrometric fit to the reference stars, in a 20$\arcsec$
circular aperture using \daophot/\textsc{phot} within
\iraf\footnote{
IRAF is distributed by the National Optical Astronomy Observatories,
which are operated by the Association of Universities for Research in
Astronomy, Inc., under cooperative agreement with the National Science
Foundation.} \citep{tody86,tody93} 
and estimated the sky using the sigma-rejected mode in an annulus
defined around each star with inner and outer radii of 33$\arcsec$ and
60$\arcsec$ respectively.

We calculated the extinction correction based on a weighted mean flux
of comparison stars and applied this correction to each of our stars. 
We iteratively selected our comparison stars by removing any that
showed unusually noisy or variable trends in their differential light
curves. Additionally, a few exposures in the beginning and very end of
our observing sequence were removed because those observations were
made through particularly thick clouds. The resulting light curve
represents the observed counts for the star corrected for extinction
using a group of 6 comparison stars within 6$\arcmin$ separation from
\hds. The \lc\ is shown on panels 3 and 4 of \figr{fulltr}.

\subsection{Observations by HATNet}

An instrument description of the wide-field HAT telescopes was given in
\citet{bakos02,bakos04}. Here we briefly recall the relevant system
parameters. A HAT instrument contains a fast focal ratio (f/1.8) 0.11m
diameter Canon lens and Peltier-cooled CCD with a front-illuminated
\ccdsize{2K}
chip having 14\micron\ pixel size. The resulting FOV is
8.2\arcdeg\ with 14\arcsec\ pixel scale. Using a
psf-broadening technique \citep{bakos04}, careful calibration
procedure, and robust differential photometry, the HAT telescopes can
achieve 3mmag precision (rms) \lcs\ at 300s resolution for bright stars
(at $I\approx8$). The HAT instruments are operated in autonomous mode,
and carry out robotic observations every clear night.

We have set up a longitude-separated, two-site network of six HAT
instruments, with the primary goal being detection of planetary
transits in front of bright stars. The two sites are FLWO, in Arizona,
the same site where the 1.2m telescope is located (\secr{flwo48obs}),
and the roof of the Submillimeter Array atop Mauna Kea, Hawaii (MK).

In addition to the wide-field HAT instruments, we developed a dedicated
photometry follow-up instrument, called TopHAT, which is installed at
FLWO\@. A brief system description was given in \citet{dc06} in context
of the photometry follow-up of the \mbox{HD 149026} planetary transit. This
telescope is 0.26m diameter, f/5 Ritchey-Cr\'etien design with a Baker
wide-field corrector. The CCD is a \ccdsize{2K} Marconi chip with
13.5\micron\ pixel size. The resulting FOV is 1.3\arcdeg\ with
2.2\arcsec\ pixel resolution. Similarly to the HATs, TopHAT is fully
automated.

Selected stations of the HAT Network, along with TopHAT, observed one
full and six partial transits of \hds\ (for details, see
\tabr{obsstat}). Observing conditions of the full-transit event at FLWO
at $N_{tr}=6$ have been summarized in \secr{flwo48obs}. This transit
was observed by HAT-5 and HAT-6 (both in I-band), and by TopHAT
(V-band). The partial transit observations at numerous later epochs
included HAT-5 (FLWO, I-band), HAT-9 (MK, I-band) and TopHAT (FLWO,
V-band). Typical exposure times for the wide-field instruments were 60
to 90 seconds with 10 second readout. TopHAT exposures were $\sim$12sec
long with up to 40 second readout and download time. All observations were
made at slight defocusing and using the psf-broadening technique. The
stellar profiles were 2.5pix (35\arcsec) and 4.5pix (9.9\arcsec) wide
for the HATs and TopHAT, respectively. Although we have no
auto-guiding, real-time astrometry was performed after the exposures,
and the telescope's position was kept constant with 20\arcsec\
accuracy.

\paragraph{Reduction and photometry}

All HAT and TopHAT images were subject to overscan correction,
two-dimensional residual bias pattern and dark subtraction, and
normalization with a master sky-flat frame. Bias, dark and sky-flat
calibration frames were taken each night by each telescope, and all
object frames were corrected with the master calibration images that
belonged to the specific observing session. Saturated pixels were
masked before the calibration procedure.

We used the 2MASS All-Sky Catalog of Point Sources
\citep{skrutskie00,cutri03} as an input astrometric catalogue, where
the quoted precision is 120 mas for bright sources. A 4th order
polynomial fit was 
used to transform the 2MASS positions to the reference frame of the
individual images. Typical rms of the transformations was 
700 mas for the wide-field instruments, and 150 mas for TopHAT. 

Fixed center aperture photometry was performed for all these stars. For
the wide-field HAT telescopes we used an $r_{ap}=3$ pixel (42\arcsec)
aperture, surrounded by an annulus with inner and outer radii of
$r_1=5$ pix (70\arcsec) and $r_2=13$ pix (3\arcmin), respectively. For
TopHAT, the best aperture was $r_{ap}=5$ pix (10.8\arcsec) with $r_1=13$
pix (29\arcsec) and $r_2=21$ pix (46\arcsec). The apertures were small
enough to exclude any bright neighboring star.

A high quality reference frame was selected for the wide-field HAT
telescopes from the Mauna Kea HAT-9 data, and separately for TopHAT\@. 
Because the HAT wide field instruments are almost identical, we were
able to use the HAT-9 reference frame to transform the instrumental
magnitudes of HAT-5 and HAT-6 data to a common system.  For this, we
used 4th order polynomials of the magnitude differences as a function
of X and Y pixel positions.
In effect, we thereby used $\sim 3000$ and $\sim 800$ selected
non-variable comparison stars for the HATs, and TopHAT, respectively.
This contributes to the achieved precision, which is only slightly
inferior to the precision achieved by the bigger diameter telescopes.

The amount of magnitude correction for \hds\ between the reference and
the individual images is shown in the $\Delta M_{ext}$ column of
\tabr{lc}. The same table also indicates the rms of these magnitude
fits in the $\sigma_{mfit}$ column. Both quantities are useful for
further cleaning of the data. Because \hds\ is a bright source, it was
saturated on a small fraction of the frames. Saturated
data-points were flagged in the \lcs, and also de-selected
from the subsequent analysis (flagged as ``C'' in \tabr{lc}).

After cleaning outliers by automatically de-selecting points where the
rms of the magnitude transformations was above a critical threshold
(typically 25mmag) the \lcs\ reached a precision of $\sim$4mmag at 90
second resolution for both the HATs and TopHAT\@. Full-transit data are
shown in panels 1, 2 and 5 of \figr{fulltr}, and partial-transit data
are shown in \figr{partr}.

\subsection{Observations by the Wise 1.0m telescope}
\label{sec:wise}

The Wise 1m f/7 telescope was used to observe the $N_{tr}=4$ and
$N_{tr}=22$ transits in B-band. The CCD was a \ccdsize{1K} Tektronics
chip with 24\micron\ pixel size that corresponds to 0.696\pxs\
resolution on the sky, and a FOV of 11.88\arcmin. The photometric
conditions were acceptable on both nights, with FWHM$\approx$2\arcsec.
Auto-guiding was used during the observations. Frames were calibrated
in a similar manner to the FLWO1.2m observations, using twilight flats,
and aperture photometry was performed with \daophot.

Unfortunately, out-of-transit (OOT) data of the first transit
($N_{tr}=4$) (which was also observed from OHP in R-band) are missing,
so it is impossible to obtain useful normalization or to apply
extinction correction to the transit curve. The second transit
($N_{tr}=22$) was processed using an aperture of 10\arcsec\ encircled
by an annulus with inner and outer radii of 15\arcsec\ and 25\arcsec,
respectively. Seven comparison stars were used, all of them bright,
isolated and far from the boundary of the FOV\@. Extinction correction,
derived from the OOT points only, was applied to the resulting stellar
\lc. The final curve of this transit is plotted in \figr{partr}.

\subsection{The resulting \lc}

All photometry originating from the individual telescopes which
contains significant OOT data has been merged, and is presented in
\tabr{lc}.  We give both the ratio of the observed flux to the OOT flux
of \hds\ (``FR''), and magnitudes that are very close to the standard
Johnson/Cousins system (``Mag'').  Due to the different observing
conditions, instruments, photometry parameters (primarily the aperture)
and various systematic effects (changing FWHM), the zero-point of the
observations were slightly offset.  Even for the same instrument,
filter-setup, and magnitude reference frame, the zero-points in the
flat OOT section were seen to differ by $0.03$mag. The offset can be
explained by long-term systematic variations and by intrinsic variation
of \hds.

In order to correct for the offsets, for each transit observation (as
indicated by $N_{tr}$ in \tabr{lc}) we calculated both the median value from
the OOT section by rejecting outliers, and also the rms around the
median.  The OOT median was used for two purposes.  First, we
normalized the flux values of the given \lc\ segment at
$N_{tr}$, which are shown in the ``FR'' (flux-ratio) column of
\tabr{lc}.  Second, we shifted the magnitudes to the standard system
in order to present reasonable values in the ``Mag'' column. For the
standard system we used the Hipparcos values, except for R-band, which
was derived by assuming $R-I = 0.48$ from \citet{cox00}.

The formal magnitude errors that are given in the ``Merr'' column are
based on the photon-noise of the source and the background-noise
\citep[e.g.~][]{newberry91}. They are in a self-consistent system, but they
underestimate the real errors, which have contributions from other noise
factors, such as i) scintillation \citep{young67, gilliland88}, ii)
calibration frames \citep{newberry91}, iii) magnitude transformations
depending on the reference stars and  imperfectly corrected extinction
(indicators of this error source are the $\Delta M_{ext}$ extinction
and the $\sigma_{mfit}$ rms of extinction corrections in \tabr{lc}). 
Because it is rather difficult to calculate these factors, we assumed
that the observed rms in the OOT section of the \lcs\ is a
relevant measure of the overall noise, and used this to normalize the
error estimates of the individual flux-ratios (column ``FRerr'', see
later \secr{mergers}).

\notetoeditor{This is the original place of \tabr{lc}. In the journal
paper we give only a short extract of the real table. In the electronic
edition the input file is \verb@tab_photdata_all.tex@}
\begin{deluxetable*}{llllllllllll}
\tabletypesize{\scriptsize}
\tablecaption{
The \lc\ of \hds.
\label{tab:lc}}
\tablewidth{0pt}
\tablehead{
	\colhead{Tel.} &
	\colhead{Fil.} &
	\colhead{$N_{tr}$} &
	\colhead{HJD} &
	\colhead{Mag} &
	\colhead{Merr} &
	\colhead{FR} &
	\colhead{$\rm FR_{corr}$} &
	\colhead{$\rm FR_{err}$} &
	\colhead{$\Delta M_{ext}$} &
	\colhead{$\sigma_{mfit}$} &
	\colhead{Qflag}\\
	\colhead{} &
	\colhead{} &
	\colhead{} &
	\colhead{} &
	\colhead{} &
	\colhead{(mag)} &
	\colhead{} &
	\colhead{} &
	\colhead{} &
	\colhead{(mag)} &
	\colhead{(mag)} &
	\colhead{}
}
\startdata
  OHP1.2 & B    & 0    & 2453629.3205430 & 8.6062 &   \nodata & 0.99614 & 0.99612 &    \nodata &   \nodata &   \nodata & \nodata \\
 FLWO1.2 & r    & 6    & 2453642.6001600 & 7.1886 & 0.0008 & 1.02934 & 1.02972 & 0.00074 &   \nodata &   \nodata & \nodata \\
   HAT-5 & I    & 6    & 2453642.5903353 & 6.7452 & 0.0017 & 0.99522 & 0.99511 & 0.00156 & -0.147 & 0.0142 &    G \\
   HAT-6 & I    & 6    & 2453642.6082715 & 6.7357 & 0.0017 & 1.00397 & 1.00406 & 0.00156 & -0.011 & 0.0160 &    G \\
  TopHAT & V    & 6    & 2453642.6042285 & 7.6717 & 0.0009 & 0.99844 & 0.99841 & 0.00083 & -0.097 & 0.0080 &    G \\
   HAT-9 & I    & 7    & 2453644.8307479 & 6.7383 & 0.0019 & 1.00157 & 1.00160 & 0.00175 & -0.015 & 0.0081 &    G \\
 Wise1.0 & B    & 22   & 2453678.1963150 & 8.6374 & 0.0013 & 0.96792 & 0.96779 & 0.00120 &   \nodata &   \nodata & \nodata \\
\enddata
\tablecomments{
This table is published in its entirety (2938 lines) in the electronic
edition of the paper. A portion is shown here regarding its form and
content with a sample line for each telescope in the order they
observed a transit with an OOT section. Column $\mathbf {N_{tr}}$ is
the number of transits since the discovery data by OHP on $\rm
HJD=2453629.3$. Values in the {\bf Mag} (magnitude) column have been
derived by shifting the zero-point of the particular dataset at
$N_{tr}$ to bring the median of the OOT section to the standard
magnitude value in the literature.  {\bf Merr} (and $\rm FR_{err}$) 
denote the {\em
formal}\/ magnitude (and flux-ratio) error estimates based on the photon and 
background noise (not available for all data).
The flux-ratio $\bf FR$ shows the ratio of the individual flux
measurements to the sigma-clipped median value of the OOT at that
particular $N_{tr}$ transit observation.
The merger-corrected flux-ratio {\bf $\rm FR_{corr}$} is described in
detail in \secr{mergers}. 
The $\mathbf{\Delta M_{ext}}$ is a measure of the extinction on a
relative scale (instrumental magnitude of reference minus image),
$\mathbf{\sigma_{mfit}}$ is the rms of the magnitude fit between the reference
and the given frame. Both of these quantities are useful measures of
the photometric conditions. {\bf Qflag} is the quality flag: ``G''
means good, ``C'' indicates that the measurement should be used with
caution, e.g.~the star was marked as saturated. Fit of the transit
parameters were performed using the HJD, FR, $\rm FR_{corr}$ and 
$\rm FR_{err}$ columns.
}
\end{deluxetable*}

\subsection{Merger analysis}
\label{sec:mergers}


\hds\ has a number of faint,
close-by neighbors that can distort the \lc, and may bias the derived
physical parameters.
These blends can have the following second-order effects:
i) the measured transit will appear shallower, as if the planetary
radius was smaller,
ii) the depth and shape of the transit will be color-dependent in a
different way than one would expect from limb-darkening models,
iii) differential extinction can yield an asymmetric \lc,
iv) variability of a faint blend can influence the observed \lc. 
Our goal was to calculate the additional flux in the various apertures
and bandpasses shown in \tabr{obsstat}, and correct our observed
flux-ratios (\tabr{lc}, column FR) to a realistic flux-ratio ($\rm
FR_{corr}$). 

The 2MASS point-source catalogue \citep{cutri03} lists some 30 stars
within 45\arcsec, which is the aperture used at the HAT-5,6,9
telescopes, 5 stars within 20\arcsec\ (FLWO1.2m), and 3 stars within
15\arcsec, which may affect the measurements of the 10\arcsec\
apertures of OHP1.2, Wise1.0 and TopHAT (apertures are listed in
\tabr{obsstat}).
To check the reality of listed blends, and to search for additional
merger stars, we inspected the following sources: the Palomar
Observatory Sky Survey (POSS I) red plates (epoch 1951), the Palomar
Quick-V Survey (QuickV, epoch 1982), the Second Palomar Sky Survey
(POSS II) plates (epoch 1990 \--- 1996), the 2MASS J, H and $K_S$ scans
(epoch 2000), and our own images.
We can use the fact that \hds\ is a high proper motion star with
velocity of $\sim 0.25\arcsec/yr$ pointing South. It was
$\sim13\arcsec$ to the North on the POSS-I plates, and $\sim 4\arcsec$
N on POSS-II, thus we can check its {\em present}\/ place when it was not
hidden by the glare of \hds. The analysis is complicated by the
saturation, diffraction spikes, and the limited scan resolution
(1.7\pxs) of POSS-I, but we can confirm that there is no significant
source at the epoch 2005 position of \hds\ down to $\sim$4mag 
fainter in R-band. The reality of all the 2MASS entries was double
checked on the POSS frames. 

There are only two additional faint sources that are missing from the
2MASS point-source catalogue, but detected by our star-extraction on
the 2MASS J, H and K scans;
the first one at $\alpha = 20^h00^m45.12^s$, 
$\delta = +22\arcdeg42\arcmin36.5\arcsec$
and the second at $\alpha = 20^h00^m43.20^s$, 
$\delta = +22\arcdeg42\arcmin42.5\arcsec$. 
We made sure these sources are not filter-glints or persistence effects
on the 2MASS scans; they are also visible on the POSS frames. Their
instrumental magnitude was transformed to the J,H,$K_S$ system using
the other stars in the field that are identified in the point source
catalogue.

A rough linear transformation was derived between the 2MASS J, H and
$K_S$ colors and Johnson/Cousins B, V, R and I by cross-identification
of $\sim 450$ \citet{landolt92} standard stars, and performing linear
regression. The uncertainty in the transformation can be as large as
0.1mag, but this is adequate for the purpose of estimating the extra
flux (in BVRI), which is only about a few percent that of \hds.

We find that the extra flux in a 45\arcsec\  aperture is $\delta=$
1.012, 1.016, 1.018 and 1.022 times the flux of \hds\ in B, V, R and
I-bands, respectively. The dominant contribution comes from the red
star 2MASS 20004297+2242342 at 11.5\arcsec\ distance, which is
$\sim$4.5mag fainter. This star has been found \citep{bakos06} 
to be a physical companion to \hds\ and thus may also be called
\hdsB\ (not to be confused with \hdsb). For the 10\arcsec\ aperture 
we assumed that half the flux of \hdsB\ is within the aperture. The
same $\delta$ flux contribution in the 10\arcsec\ aperture is 1.003,
1.005, 1.006 and 1.008 in BVRI\@. The corrected flux-ratios of the
individual measurements to the median of the OOT were calculated in the
manner $\rm FR_{corr} = 1 + \delta \cdot (FR - 1)$, and are shown in
\tabr{lc}. There is a small difference ($\sim$2\%) between the
10\arcsec\ and 45\arcsec\ flux contribution, thus we expect that the
former measurements (OHP1.2, Wise1.0, TopHAT) show slightly deeper
transits than the FLWO1.2m (r) and wide-field HATNet telescopes (I).

\section{Deriving the physical parameters of the system}
\label{sec:physpar}

We use the full analytic formula for nonlinear limb-darkening given in
\citet{mandel02} to calculate our transit curves.  
In addition to the orbital period and limb-darkening coefficients, 
these curves are a
function of four variables, including the mass (\mstar) and radius
(\rstar) of the star, the radius of the planet (\rpl), and the
inclination of the planet's orbit relative to the observer (\ipl).
Because these parameters are degenerate in the transit curve, we use
$\mstar=0.82\pm0.03\msun$ from \citetalias{bouchy05} to break the
degeneracy. As regards \rstar, there are two possible approaches:
i) assume a fixed value from independent measurements (\secr{stelrad}), 
ii) measure the radius of the star directly from the transit curve,
i.e.~leave it to vary freely in the fit. Our final results are based on
detailed analysis (\secr{trfit}) using the first approach. To fully
trust the second approach, one would need high precision data with
relatively small systematic errors. Nevertheless, in order to check
consistency, we also performed an analysis where the stellar radius was
left as a variable in the fit, and also checked the effect of
systematic variations in the \lcs\ (see later in \secr{trfit}). Refined
ephemerides and center of transit time residuals are discussed in
\secr{ephem}.

\subsection{The radius of \hds}
\label{sec:stelrad}

Because the value of the stellar radius we use in our fit linearly
affects the size of the planet radius we obtain, we use several
independent methods to check its value and uncertainty.

\paragraph{First method}
For our first calculation we use 2MASS \citep{cutri03} and Hipparcos
photometry \citep{perryman97} to find the V-band magnitude and V-K
colors of the star, and use the relation described in
\citet{kervella04} to find the angular size of the star.  Because this
relation was derived using Johnson magnitudes, we first convert the
2MASS $\Ks = 5.541\pm 0.021$ magnitude to the Bessell-Brett homogenized
system, which in turn is based on the SAAO system, and thus is the
closest to Johnson magnitude available \citep{carpenter05}.  We obtain
a value of $K = 5.59\pm0.05$. Most of the error comes from the
uncertainty in the $\rm J-\Ks$ color, which is used in the conversion.

The Johnson V-band magnitude from Hipparcos is $7.67\pm0.01$.  This
gives a V-K color of $2.09\pm0.06$.  From \citet{kervella04}, the
limb-darkened angular size of a dwarf star is related to its K
magnitude and V-K color by:
\begin{equation}\label{eq:kervella}
log(\theta)=0.0755(V-K)+0.5170-0.2K\,.
\end{equation}
Given the proximity of \hds\ ($19.3\pm0.3$pc), reddening can be
neglected, despite its low galactic latitude. The relation gives an
angular size of $0.36\pm0.02$mas for the stellar photosphere, where
the error estimate originates from the errors of V-K and K. The small
dispersion of the relation was not taken into account in the error
estimate, as it was determined by \citet{kervella04} using a fit to a
sample of stars with known angular diameters to be less than 1\%. 
Using the Hipparcos parallax we find that $\rstar = 0.75\pm0.05 \rsun$.

\paragraph{Second method}
We  also derive the  radius of  the star  directly from  the Hipparcos
parallax, V-band magnitude, and temperature of the star.  We first
convert from apparent magnitude to absolute magnitude and apply a
bolometric correction \citep{bessel98}. To solve for the radius of the
star we use the relation:
\begin{equation}
M_b=4.74-2.5log\left[\left(\frac{\teffstar}{\teffsun}\right)^4
	\left(\frac{\rstar}{\rsun}\right)^2\right]
\end{equation}
For $\teffstar=5050\pm50$ K effective temperature \citepalias{bouchy05}, we
measure a radius of $0.74\pm0.03 \rsun$.

\paragraph{Third method \--- isochrones}
An additional test on the stellar radius and its uncertainty comes from
stellar evolution models. We find from the \citet{girardi02} models
that the isochrone gridpoints in the (\teff, \logg) plane that are
closest to the observed values ($\teffstar=5050\pm50K$,
$\logg=4.53\pm0.14$) prefer slightly evolved models with
$\mstar\approx0.80\msun$ and $\rstar\approx0.79\rsun$. Alternatively,
the Hipparcos $V=7.67\pm0.01$ magnitude, combined with the 
$m-M = -1.423 \pm 0.035$ distance modulus yields $M_V = 6.25\pm0.04$ 
absolute V magnitude, and the closest isochrone gridpoints prefer less
evolved stars with $\mstar\approx0.80\msun$ and
$\rstar\approx0.76\rsun$. The discrepancy between the above two approaches
decreases if we adopt a slightly larger distance modulus. Finally,
comparison to the \citet{baraffe98} isochrones yields
$\mstar\approx0.80\msun$ and
$\rstar\approx0.76\rsun$. From isochrone fitting, the error on the
stellar radius can be as large as 0.03\rsun.

\paragraph{Fourth method}
Recently \citet{masana06} calibrated the effective temperatures,
angular semi-diameters and bolometric corrections for F, G, K type
stars based on V and 2MASS infrared photometry. They provide \--- among
other parameters \--- angular semi-diameters and radii for a large
sample of Hipparcos stars. For \hds\ they derived $R =
0.758\pm0.016\rsun$.

\paragraph{Summary}
Altogether, the various methods point to a stellar radius in the range
of 0.74 to 0.79\rsun, with mean value being $\sim 0.76\rsun$. In the
subsequent analysis we accept the \citet{masana06} value of
$0.758\pm0.016$\rsun.

\subsection{Fitting the transit curve}
\label{sec:trfit}

We set the mass and radius of the star equal to $0.82\pm0.03\msun$ 
and $0.758\pm0.016$ $\rsun$, respectively, and fit for the planet's
radius and orbital inclination.  The goodness-of-fit parameter is given
by:
\begin{equation}
\label{eq:xi2}
\chi^2=\sum_{i=1}^N\left(\frac{p_i-m_i}{\sigma_{m,i}}\right)^2
\end{equation}
where $m_i$ is the $\textrm{i}^{th}$ measured value for the flux from
the star (with the median of the out of transit points normalized to
one), $p_i$ is the predicted value for the flux from the theoretical
transit curve, and $\sigma_{m,i}$ is the error for each flux measurement.

For the OHP1.2 and Wise data, where independent errors for each flux
measurement are not available, we set the $\sigma_{m,i}$ errors on all
points equal to the standard deviation of the out of transit points.
For the FLWO1.2, HAT and TopHAT data, where relative errors for
individual points are available, we set the median error equal to the
standard deviation of the out of transit points and use that to
normalize the relative errors. We also allow the locations of
individual transits to vary freely in the fit.

When calculating our  transit  curves, we  use the  nonlinear
limb-darkening law defined in \citet{claret00}:
\begin{equation}
I(r)=1-\sum_{n=1}^4c_n(1-\mu^{n/2})
\end{equation}
where
\begin{equation}
\mu=cos\theta\,.
\end{equation}
We select the four-parameter nonlinear limb-darkening coefficients from
\citet{claret00} for a star with $T=5000K$, $log(g)=4.5$,
$[Fe/H]=0.0$, and a turbulent velocity of 1.0 km/s.  The actual
parameters for the star, from \citetalias{bouchy05}, are rather close
to this: $T=5050\pm50K$, $log(g)=4.53\pm0.14$, and $[Fe/H]=-0.03\pm0.04$.

\begin{figure}
\plotone{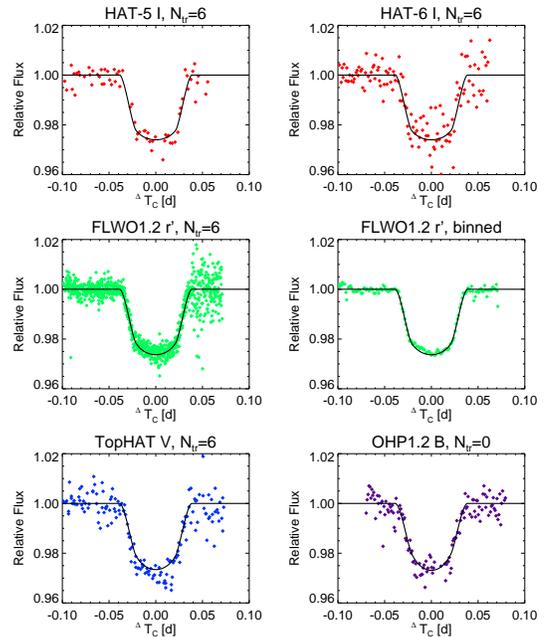}
\caption{
The five full eclipses examined in this work, with best-fit
transit curves over-plotted. The figure in the electronic edition is
color-coded according to the bandpass used.
\label{fig:fulltr}}
\end{figure}

\notetoeditor{This is the original place of \figr{fulltr}. Color version
is to be placed in electronic edition only. B\&W version also included.}

\notetoeditor{This is the original place of \tabr{best_fit_coeff}.}
\begin{deluxetable}{lrrrrcrrrrr}
\tabletypesize{\scriptsize}
\tablecaption{
Parameters from simultaneous fit of transit
curves.
\label{tab:best_fit_coeff}}
\tablewidth{0pt}
\tablehead{
	\colhead{Parameter} &
	\colhead{Best-Fit Value}
}
\startdata
\rpl (\rjup) & 1.154 $\pm0.032$ \\
\ipl (\arcdeg) & 85.79 $\pm0.24$ \\
\mstar (\msun) & 0.82 $\pm0.03$\tablenotemark{a}\\
\rstar (\rsun) & 0.758 $\pm0.016$\tablenotemark{b}\\
Period (days) & $2.218573\pm0.000020$\\
$T_0$ (HJD) & $2453629.39420\pm0.00024$\\
\enddata
\tablenotetext{a}{From \citetalias{bouchy05}}
\tablenotetext{b}{From \citet{masana06}}
\end{deluxetable}

To determine the  best-fit radius  for  the planet,  we evaluate  the
$\chi^2$ function over all {\em full}\/ transits simultaneously, using
the same values for the planetary radius and inclination. For this
purpose, we employed the downhill simplex minimization routine
(\textsc{amoeba}) from \citet{press92}. The full transits and the
fitted curves are exhibited on \figr{fulltr}, the transit parameters
are listed in \tabr{best_fit_coeff}. In order to determine the
$1\sigma$ errors, we fit for the inclination and the radius of the
planet using the $1\sigma$ values for the mass and radius of the star 
(assuming they are uncorrelated).  We find that the mass of the star
contributes errors of
$\pm0.004$ $\rjup$ and $\pm0.12$\arcdeg, 
and the radius of the star contributes errors of
$\pm0.032$ $\rjup$ and $\pm0.21$\arcdeg.  
Using a bootstrap Monte Carlo method, we also estimate the errors from
the scatter in our data, and find that this scatter contributes an
error of $\pm0.005$ $\rjup$ and $\pm0.03$\arcdeg\ to the final
measurement.  This gives us a total error of
$\pm0.032$ $\rjup$ for the planet radius and $\pm0.24$\arcdeg\ for the 
inclination.

Our best-fit parameters gave a reduced $\chi^2$ value of 1.23.  The
excess in the reduced $\chi^2$ over unity is the result of our method
for normalizing the relative errors for data taken at $N_{tr}=6$, where
the RMS variation in the data increases significantly towards the end
of the data set, as the source moved closer to the horizon.  For these
data we define our errors as the standard deviation of the data before
the transit, where the scatter was much smaller. This is justified
because we know from several sources (night webcamera, raw photon
counts) that the conditions were similar (photometric) before and
during the transit, and the errors before the transit better represent
those inside the transit. This underestimates the errors for data after
the transit, inflating the $\chi^2$ function accordingly. We find that
when we exclude the FLWO1.2 data after the end of the transit (the
FLWO1.2 data contain significantly more points than any other single
data set), the reduced $\chi^2$ for the fit decreases to 0.93.

The results  of  the  planet  transit  fit  are  shown  in
\tabr{best_fit_coeff}. The value for the radius of the planet
$\rpl=1.15\pm0.03$ $\rjup$ is smaller than the
\citetalias{bouchy05} value 
($\rpl=1.26\pm0.03$ $\rjup$), and the inclination of 
$85.8\pm0.2$\arcdeg\ is slightly larger than the \citetalias{bouchy05} 
value
($85.3\pm0.1$\arcdeg). Although our errors are comparable to the errors
given by \citetalias{bouchy05}, despite the superior quality of the new
data, we note that this is a direct result of
the larger error ($\pm0.016$ instead of $\pm0.01$) for the stellar
radius we use in our fits.  As discussed in \secr{stelrad}, we feel
that this error, which is based on the effective temperature and
bolometric magnitude of the star, is a more accurate reflection of the
uncertainties in the measurement of the radius of the star. 
We note that the errors are dominated by the uncertainties in the
stellar parameters (notably \rstar).

\paragraph{Fitting with unconstrained stellar radius}

We note that when we fit for the stellar radius directly from
the transit curves (meaning we fit for the planet radius, orbital
inclination, and stellar radius, but set the stellar mass equal to
$0.82$ $\msun$), we measure a stellar radius of $0.678\pm0.015$
$\rsun$ and planet radius of $\rpl = 0.999\pm0.026\rjup$.  The errors 
for these measurements are from a bootstrap Monte Carlo analysis, and
represent the uncertainties in our data alone.  To obtain the formal
errors, we incorporate the error from the mass of the star and find
errors of $\pm0.017\rsun$ and $0.029\rjup$, respectively.  This means
that our data prefer a significantly smaller stellar radius (and a
correspondingly smaller planet radius) than our estimates based on
temperature, bolometric magnitude, and V-K colors alone would lead us
to expect, or a radius smaller than the 0.82\msun\ stellar mass implies.

With many more points (869 as compared to $\sim 100$ in the other
data-sets) and lower photon-noise uncertainties, the FLWO1.2 data
dominate the fit of \eqr{xi2}. However, we repeated our fit with and
without these data, and found that the best-fit radius for the star
decreased only slightly (to 0.666 \rsun) when the FLWO1.2 data were
excluded from the fit.
Thus, our I, V, and B-band data independently yield values for the
stellar radius similar to those implied by the FLWO1.2 data.

\paragraph{The effect of systematic errors}

The $\chi^2$ minimization formula (\eqr{xi2}) assumes independent
noise, and the presence of covariance in the data (due to systematics
in the photometry) means that too much weight may be given to a dataset
having small formal errors and a great number of datapoints (e.g.~the
FLWO1.2 data) compared to the other independent data sets (e.g.~other
telescopes and filters). This is especially a concern when the datasets
yield different transit parameters, and one needs to establish whether
this difference is significant. 
In order to follow-up this issue, we repeated the global fit by
assuming that the photometric systematics were dominant in the error
budget on the parameters \--- as suggested by our experience with
milli-magnitude rapid time-series photometry. We estimated the
amplitude of the covariance from the variance of 20-minute sliding
averages on the residuals around the best-fitting transit \lc\ for each
night following the method of \citet{pont05b}. The fit was repeated
using these new weights (listed in \tabr{obsstat}, $\sigma_{sys}$), and
the resulting parameters (planetary radius, inclination) were within
1\% of the values found assuming independent noise. The dispersion of
these parameters from the individual nights were found to be compatible
with the uncertainties due to the systematics. The amplitude of the
systematics is also sufficient to account for the difference in the
best-fit stellar radius if it is left as a free parameter. Therefore,
with the amplitude of the covariance in the photometry determined from
the data itself, we find that the indications of discrepancy between
the different data sets and with the assumed primary radius are not
compelling at this point. 

\subsection{Ephemerides}
\label{sec:ephem}

\begin{figure}
\plotone{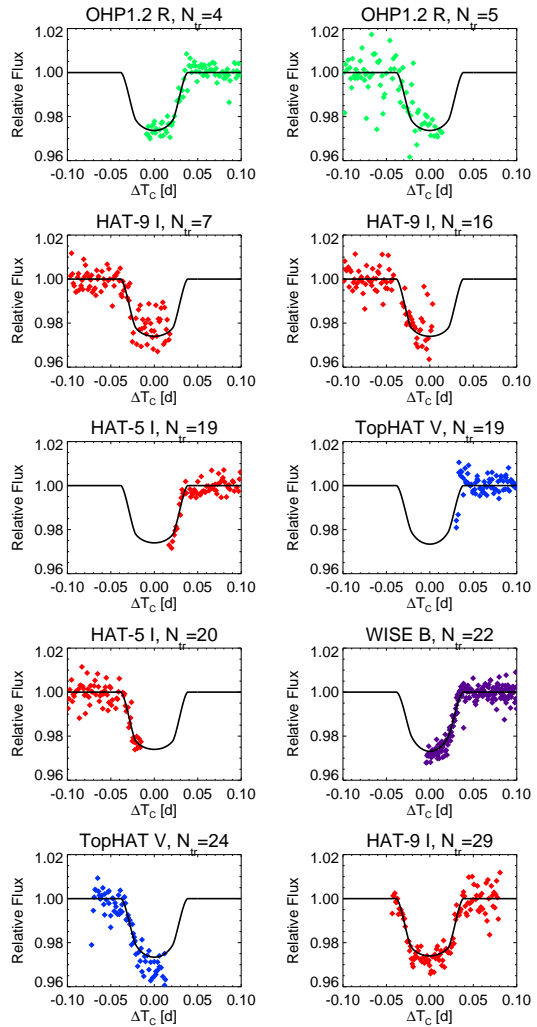}
\caption{
The ten partial eclipses examined in this work, with
best-fit transit curves over-plotted. The eclipses are listed
sequentially by date, from top left to bottom right. These eclipses
were not used in the fit for the planet radius, inclination, stellar
mass, and stellar radius. The figure in the electronic edition is
color-coded according to the bandpass used.
\label{fig:partr}}
\end{figure}

\notetoeditor{This is the original place of \figr{partr}. Color version
is to be placed in electronic edition only. B\&W version also included.}

\notetoeditor{This is the original place of \tabr{ocres}.}
\begin{deluxetable}{lrrrrcrrrrr}
\tabletypesize{\scriptsize}
\tablecaption{Best-Fit Transit Locations\label{tab:trloc}}
\tablewidth{0pt}
\tablehead{
	\colhead{Telescope} & 
	\colhead{$N_T$} & 
	\colhead{$T_C$} & 
	\colhead{$\sigma_{\textrm{HJD}}$} & 
	\colhead{(O-C)} &
	\colhead{$\frac{\textrm{(O-C)}}{\sigma_{\textrm{HJD}}}$}\\
	\colhead{} & 
	\colhead{} & 
	\colhead{(HJD, days)} & 
	\colhead{(days)} & 
	\colhead{(days)} &
	\colhead{}
}
\startdata
OHP1.2 & 0 & 2453629.39073  & $\pm0.00059$ & $-0.0035$ & $-5.9$\\
OHP1.2 & 4 & 2453638.26885  & $\pm0.00067$ & 0.00035   & 0.53\\
OHP1.2 & 5 & 2453640.48706  & $\pm0.00174$ & $-0.0000079$ & $-0.0045$\\
FLWO1.2 & 6 & 2453642.70592 & $\pm0.00022$ & 0.00029   & 1.24\\
HAT-5 & 6 & 2453642.70641   & $\pm0.00092$ & 0.00077   & 0.84\\
HAT-6 & 6 & 2453642.70649   & $\pm0.00049$ & 0.00085   & 1.7\\
TopHAT & 6 & 2453642.70536  & $\pm0.00048$ & $-0.00028$ & $-0.57$\\
HAT-9 & 7 & 2453644.92720   & $\pm0.00111$ & 0.0030    & 2.7\\
HAT-9 & 16 & 2453664.89287  & $\pm0.00108$ & 0.0015    & 1.4\\
HAT-5 & 19 & 2453671.54999  & $\pm0.00113$ & 0.0029    & 2.6\\
TopHAT & 19 & 2453671.54849 & $\pm0.00096$ & 0.0014    & 1.5\\
HAT-5 & 20 & 2453673.76725  & $\pm0.00072$ & 0.0016    & 2.2\\
Wise & 22 & 2453678.20080   & $\pm0.00050$ & $-0.0020$ & $-4.0$\\
TopHAT & 24 & 2453682.63715 & $\pm0.00100$ & $-0.0028$ & $-2.8$\\
HAT-9 & 29 & 2453693.73327  & $\pm0.00090$ & 0.00045   & 0.51\\
\enddata
\tablecomments{These are the best-fit locations for the centers of the
fifteen full and partial eclipses examined in this work. We also give
the number of elapsed transits $N_T$ and O-C residuals for each
eclipse.}
\end{deluxetable}

The transit curves derived from the full-transits for each bandpass
were used in turn to calculate the ephemerides of \hds\ using {\em
all}\/ transits that have significant OOT and in-transit sections
present (for reference, see \tabr{obsstat}).
For each transit (full and partial), the center of transit $T_C$ was
determined by $\chi^2$ minimization. Partial transits with the fitted
curve overlaid are exhibited on \figr{partr}. Errors were assigned to
the $T_C$ values by perturbing $T_C$ so that $\chi^2$ increases by
unity. The individual $T_C$ transit locations and their respective
errors are listed in \tabr{trloc}. The typical timing errors were
formally of the order of 1 minute. This, however, does not take into
account systematics in the shape of the light-curves. 
The errors in $T_C$ can be estimated from the simultaneous transit
observations, for example the $N_{tr} = 6$ event was observed by the
FLWO1.2m, HAT-5, HAT-6 and TopHAT telescopes (\tabr{obsstat}) and the
rms of $T_C$ around the median is $\sim50$ seconds, which is in harmony
from the above independent estimate of 1 minute.
We applied an
error weighted least square minimization on the $T_C = P\cdot N_{tr} + E$
equation, where the free parameters were the period $P$ and epoch $E$.
The refined ephemeris values are listed in \tabr{best_fit_coeff}.
They are consistent with both those derived by \citetalias{bouchy05}
and by \citet{hebrard06} using
Hipparcos and OHP1.2 data, to within $1\sigma$ using our error bars.

\begin{figure}
\plotone{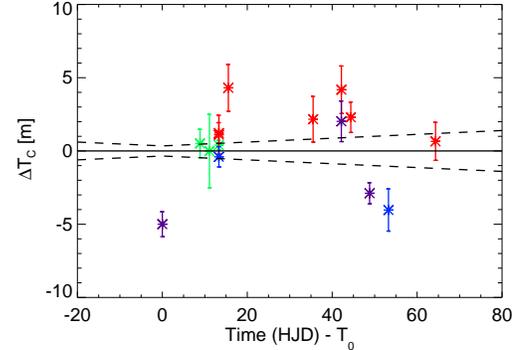}
\caption{
The residuals calculated using the period and $T_0$
derived in this work. The dashed lines are calculated from the
uncertainties in the measurements of P and $T_0$. The figure in the
electronic edition is color-coded according to the bandpass used.
\label{fig:ocres}}
\end{figure}

We also examined the Observed minus Calculated (O-C) residuals, as
their deviation can potentially reveal the presence of moons or
additional planetary companions \citep{holman05,agol05}. The O-C values
are listed in \tabr{best_fit_coeff}, and plotted in \figr{ocres}. 
Using the approximate formula from \citet{holman05}, as an example, a
0.15\mjup\ perturbing planet on a circular orbit, at 2 times the
distance of \hdsb\ ($P\approx 6.3^d$) would cause variations in the
transit timings of 2.5 minutes. The radial velocity semi-amplitude of
\hds\ as induced by this hypothetical planet would be 19m/s, which
would be barely noticeable (at the $1\sigma$ level) from the discovery
data having residuals of 15m/s and spanning only 30 days. 

A few, seemingly significant outlier points on the O-C diagram are
visible, but we believe that it would be premature to draw any
conclusions, because: i) the error-bars do not reflect the effect of
systematics, and for example the $T_C$ of the $N_{tr}=0$ OHP discovery
data moved by $\sim5$ minutes after re-calibration of that dataset, ii)
all negative O-C outliers are B or V-band data, which is
suggestive of an effect of remaining color-dependent systematics.
The significance of a few outliers is further diminished by the short
dataset we have; no periodicity can be claimed by observing 2 full and
9 partial transits altogether. 

According to the
theory, the nature of perturbations would be such that they appear as
occasional, large outliers. Thus, the detection of potential
perturbations also benefits from the study of numerous sequential
transits, for example, the MOST mission with continuous coverage and
uniform data would be suitable for such study \citep{walker03}. We also
draw the attention to the importance of observing {\em full}\/ transits, as
they improve the $T_C$ center of transit by a significant factor,
partly because of the presence of ingress and egress, and also due to a
better treatment of the systematics.

If a planet is perturbed by another planet, the
transit-time variations $\Delta t$ are proportional to the period $P$
of the perturbed planet \citep{holman05}. 
Although \hdsb\ is a relatively short period
($2.21^d$) planet compared to e.g.~\hdstb\ ($3.5^d$), it is a promising
target for detecting transit perturbations in the future, because the
mass of the host star is low and $\Delta t \propto 1/\mstar$, plus the
deep transit of the bright source will result in very precise timing
measurements. Observations spanning several months to many years may be
needed to say anything definite about the presence or absence of a
periodic perturbation.

\notetoeditor{This is the original place of \figr{ocres}. Color version
is to be placed in electronic edition only. B\&W version also included.}

\section{Conclusions}
\label{sec:conc}

Our final values for the planet radius and orbital inclination were
derived by fixing the stellar radius and mass to independently
determined values from \citetalias{bouchy05} and \citet{masana06}. We
analyzed the dataset
in two ways: by $\chi^2$ minimization assuming independent errors, and
also by assuming that photometric systematics were dominant in the
error budget. Both methods yielded the same transit parameters within
1\%:
if we assume there is no additional unresolved close-in stellar
companion to \hds\ to make the transits shallower, then we find
a planet radius of $1.154\pm0.032\rjup$ and an orbital inclination of
$85.79\pm0.24\arcdeg$ (\tabr{best_fit_coeff}). The
uncertainty in \rpl\ is primarily due to the uncertainty in the
value of the stellar radius.

\begin{figure}
\plotone{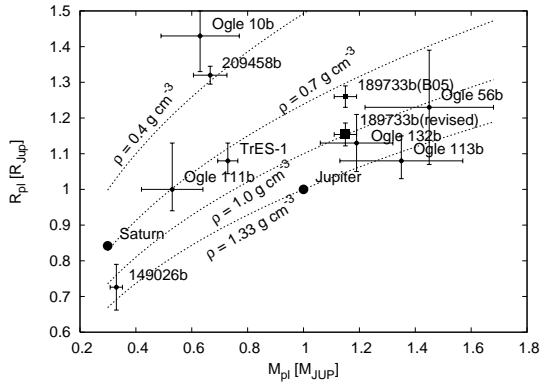}
\caption{
Mass-radius diagram depicting known transiting exoplanets,
 plus Saturn and Jupiter (for comparison). Both the discovery
(\citetalias{bouchy05}, small filled square) and revised radius (large
filled square) for \hds\ is shown. Sources
for the mass and radius values (in this order) are listed after the
name of the planet; 
\mbox{HD 209458b}: \citet{laughlin05}, Knutson et.~al.~2006, in
preparation, 
\hds: \citetalias{bouchy05}; this paper, 
\mbox{OGLE 111b}: \citet{pont04}, 
\mbox{OGLE 10b}: \citet{konacki05}, \citet{santos06},
\mbox{OGLE 132b}:  \citet{moutou04},
\mbox{OGLE 56b}: \citet{torres04},
\mbox{OGLE 113b}: \citet{bouchy04}
\mbox{TrES-1}: \citet{laughlin05}
\mbox{HD 149026}: \citet{sato05}, \citet{dc06}.
\label{fig:exo_m_r}}
\end{figure}

\notetoeditor{This is the original place of \figr{exo_m_r}.}

We note that the TopHAT V-band full and partial transit data, as well
as the Wise partial B-band data appear slightly deeper than the best
fit to the analytic model. The precision of the dataset is not adequate
to determine if this potential discrepancy is caused by a real physical 
effect (such as a second stellar companion) or to draw further conclusions. 

When compared to the discovery data, the radius decreased by 10\%, and
\hdsb\ is in the mass and radius range of ``normal'' exoplanets
(\figr{exo_m_r}). The revised radius estimate is consistent
with structural models of hot Jupiters that include the effects
of stellar insolation, and hence it does not require the presence
of an additional energy source, as is the case for \hdstb.
On the mass\---radius diagram \hdstb\ remains an outlier with
anomalously low density. We note that the parameters of OGLE-10b are
still debated \citep{konacki05,holman06,santos06}, 
but according to the recent analysis of
\citet{santos06}, it also has anomalously low density. 
With its revised parameters, \hdsb\ is quite
similar to OGLE-TR-132b \citep{moutou04}. The smaller radius leads to
a higher density of $\sim 1\gmc$ as compared to the former $\sim
0.75\gmc$. 
The smaller planetary radius increases the 16\micron\ brightness
temperature $T^{(16\mu)} = 1117\pm42\rm K$ of \citet{deming06} to
$1279\pm90$K, which is slightly larger than that of TrES-1 and \hdstb.

We also derived new ephemerides, and investigated the outlier points in
the O-C diagram.  We have not found any compelling evidence for
outliers that could be due to perturbations from a second planet in the
system. We note
however, that due to the proximity and brightness of the parent
star, as well as the deep transit, the system is well suited for
follow-on observations.

\acknowledgments
\acknowledgments
Part of this work was funded by NASA grant NNG04GN74G. 
Work by G.~\'A.~B.\ was supported by NASA through grant
HST-HF-01170.01-A Hubble Fellowship.
H.~K.~is supported by a National Science Foundation
Graduate Research Fellowship.
D.~W.~L.~thanks the Kepler mission for support through NASA
Cooperative Agreement NCC2-1390. 
A.~P.~wishes to acknowledge the hospitality of the Harvard-Smithsonian
Center for Astrophysics, where part of this work has been carried out.
Work of A.~P.~was also supported by Hungarian OTKA grant T-038437.
Research of T.~M.~and A.~S.~was partially supported by the German-Israeli
Foundation for Scientific Research and Development. 
This publication makes  use of data products from the Two Micron All
Sky Survey (2MASS).
We thank M.~Hicken and R.~Kirshner for swapping
nights on the FLWO1.2m telescope on short notice.




\end{document}